\documentclass[prc, reprint, amsmath, amssymb, superscriptaddress, nofootinbib, showpacs]{revtex4-1}

\usepackage{graphicx}%
\usepackage{multirow}
\usepackage{float}
\usepackage{epstopdf}
\usepackage[utf8]{inputenc}
\usepackage[english]{babel}
\usepackage{footnote}

\begin{document}

\title{Investigation of potential ultra-low $Q$-value $\beta$-decay candidates $^{89}$Sr and $^{139}$Ba using Penning trap mass spectrometry}%

\author{R. Sandler}%
\email{sandler@nscl.msu.edu}
\affiliation{Department of Physics, Central Michigan University, Mount Pleasant, Michigan, 48859, USA}
\affiliation{National Superconducting Cyclotron Laboratory, East Lansing, Michigan, 48824, USA}
\author{G. Bollen}
\affiliation{Facility for Rare Isotope Beams, East Lansing, Michigan, 48824, USA}
\affiliation{Department of Physics and Astronomy, Michigan State University, East Lansing, Michigan 48824, USA}
\author{N. D. Gamage}
\affiliation{Department of Physics, Central Michigan University, Mount Pleasant, Michigan, 48859, USA}
\author{A. Hamaker}
\affiliation{National Superconducting Cyclotron Laboratory, East Lansing, Michigan, 48824, USA}
\affiliation{Department of Physics and Astronomy, Michigan State University, East Lansing, Michigan 48824, USA}
\author{C. Izzo}
\affiliation{National Superconducting Cyclotron Laboratory, East Lansing, Michigan, 48824, USA}
\affiliation{Department of Physics and Astronomy, Michigan State University, East Lansing, Michigan 48824, USA}
\author{D. Puentes}
\affiliation{National Superconducting Cyclotron Laboratory, East Lansing, Michigan, 48824, USA}
\affiliation{Department of Physics and Astronomy, Michigan State University, East Lansing, Michigan 48824, USA}
\author{M. Redshaw}
\affiliation{Department of Physics, Central Michigan University, Mount Pleasant, Michigan, 48859, USA}
\affiliation{National Superconducting Cyclotron Laboratory, East Lansing, Michigan, 48824, USA}
\author{R. Ringle}
\affiliation{National Superconducting Cyclotron Laboratory, East Lansing, Michigan, 48824, USA}
\author{I. Yandow}
\affiliation{National Superconducting Cyclotron Laboratory, East Lansing, Michigan, 48824, USA}
\affiliation{Department of Physics and Astronomy, Michigan State University, East Lansing, Michigan 48824, USA}
\date{\today}%

\begin{abstract}
\noindent
\begin{description}
\item[Background]
Ultra-low $Q$-value $\beta$-decays are interesting processes to study with potential applications to nuclear $\beta$-decay theory and neutrino physics. While a number of potential ultra-low $Q$-value $\beta$-decay candidates exist, improved mass measurements are necessary to determine which of these are energetically allowed.
\item[Purpose]
To perform precise atomic mass measurements of $^{89}$Y and $^{139}$La. Use these new measurements along with the precisely known atomic masses of $^{89}$Sr and $^{139}$Ba and nuclear energy level data for $^{89}$Y and $^{139}$La to determine if there could be an ultra-low $Q$-value decay branch in the $\beta$-decay of $^{89}$Sr $\rightarrow$ $^{89}$Y or $^{139}$Ba $\rightarrow$ $^{139}$La.
\item[Method]
High-precision Penning trap mass spectrometry was used to determine the atomic mass of $^{89}$Y and $^{139}$La, from which $\beta$-decay $Q$-values for $^{89}$Sr and $^{139}$Ba were obtained. 
\item[Results]
The $^{89}$Sr $\rightarrow$ $^{89}$Y and $^{139}$Ba $\rightarrow$ $^{139}$La $\beta$-decay $Q$-values were measured to be $Q_{\rm{Sr}}$ = 1502.20(0.35) keV and $Q_{\rm{Ba}}$ = 2308.37(0.68) keV. These results were compared to energies of excited states in $^{89}$Y at 1507.4(0.1) keV, and in $^{139}$La at 2310(19) keV and 2313(1) keV to determine $Q$-values of -5.20(0.37) keV for the potential ultra-low $\beta$-decay branch of $^{89}$Sr and -1.6(19.0) keV and -4.6(1.2) keV for those of $^{139}$Ba.
\item[Conclusion]
The potential ultra-low $Q$-value decay branch of $^{89}$Sr to the $^{89}$Y (3/2$^-$, 1507.4 keV) state is energetically forbidden and has been ruled out. The potential ultra-low $Q$-value decay branch of $^{139}$Ba to the 2313 keV state in $^{139}$La with unknown J$^{\pi}$ has also been ruled out at the 4$\sigma$ level, while more precise energy level data is needed for the $^{139}$La (1/2$^+$, 2310 keV) state to determine if an ultra-low $Q$-value $\beta$-decay branch to this state is energetically allowed.
\end{description}
\end{abstract}

\maketitle

\section{Introduction}

Ultra-low $Q$-value $\beta$-decays, in which the parent nucleus decays to an excited state in the daughter with a $Q$-value of less than 1 keV, provide a powerful tool to test the role of atomic interference effects in nuclear $\beta$-decay~\cite{mus2010,Suh2010}. They can also potentially be used as new candidates for direct neutrino mass determination experiments~\cite{cat2005,cat2007,Suh2014,kop2010}. In order for a potential ultra-low $Q$-value decay to be identified or ruled out, precise measurements of the ground-state to ground-state $Q$-value as well as the excited state energy levels of the daughter nucleus are necessary. 

Currently, the only known ultra-low $Q$-value $\beta$-decay is that of $^{115}$In to the 3/2$^{+}$ first excited state in $^{115}$Sn. This decay branch was discovered by Cattadori, \textit{et al.} in 2005 via the observation of a 497.48 keV line in a $\gamma$-ray spectroscopy measurement on an $\sim$1 kg metallic indium sample at Gran Sasso underground laboratory~\cite{cat2005}.  Cattadori, \textit{et al.} inferred that $^{115}$In must undergo a weak $\beta$-decay branch to the 3/2$^{+}$ level in $^{115}$Sn at 497.334(22) keV\footnote{The energy of the $^{115}$Sn (3/2$^{+}$) state was recently measured more precisely to be 497.342(3) keV~\cite{Zhe2019}.}~\cite{Bla2005}. Using the atomic mass data available at the time~\cite{AME2003}, the $Q$-value was determined to be 2(4) keV. Later, Penning trap measurements of the $^{115}$In -- $^{115}$Sn mass difference performed with JYFLTRAP at Jyv{\"a}skyl{\"a} and with the MIT/FSU trap at Florida State University, combined with the daughter state energy, confirmed that this decay is energetically allowed. The JYFLTRAP and FSU groups determined the $Q$-value of the ultra-low decay branch to be 0.35(0.17) keV~\cite{wie2009} and 0.155(24) keV~\cite{mou2009}, respectively, making this the lowest known $Q$-value $\beta$-decay. The observation of the $^{115}$In $\rightarrow$ $^{115}$Sn (3/2$^{+}$) decay was later confirmed in measurements with an $\sim$2.5 kg indium sample at the HADES underground laboratory~\cite{wie2009,And2011}. However, theoretical calculations of the partial half-life for the $^{115}$In ultra-low $Q$-value decay that used the Penning trap $Q$-values showed a significant discrepancy with the experimental results~\cite{mus2010,Suh2010}. Hence, experimental data for additional ultra-low $Q$-value decays are called for.

Since the discovery of the ultra-low $Q$-value $\beta$-decay of $^{115}$In, other potential ultra-low $Q$-value decay branches were identified in $^{115}$Cd~\cite{Haa2013}, $^{135}$Cs~\cite{Mus2011}, and a number of other isotopes ~\cite{mus2010_2,Suh2014,kop2010,gam2019}. However, in all of the identified cases, more precise atomic mass data is required for the parent and/or daughter isotope. In Ref.~\cite{gam2019}, four cases were identified for which the daughter is a stable isotope whose mass is known less precisely than that of the parent. In this work, we investigate two of those systems: the decay of $^{89}$Sr $\rightarrow ^{89}$Y and $^{139}$Ba $\rightarrow ^{139}$La. 

\begin{figure}[h]
        \includegraphics[width=0.85\columnwidth]{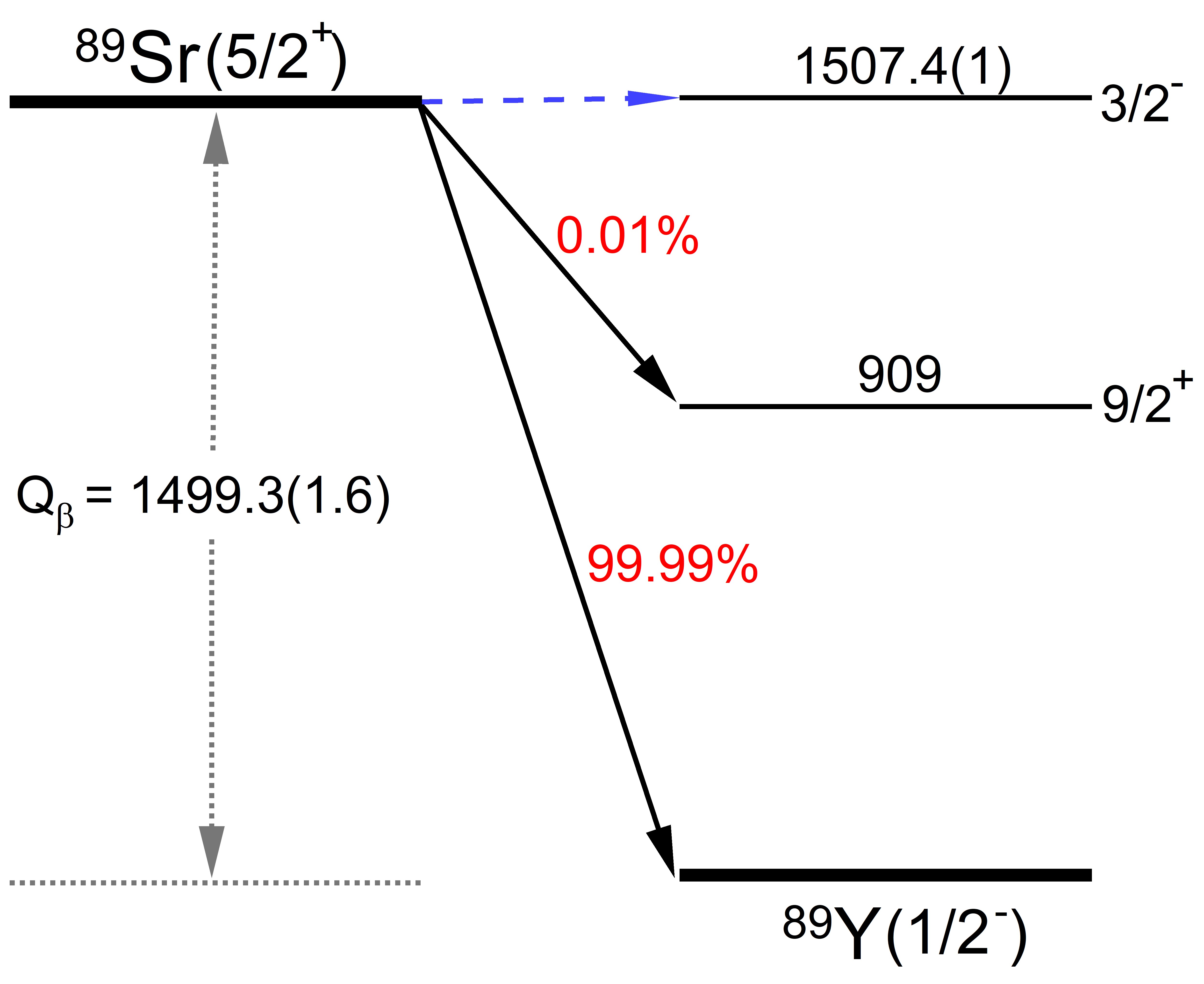}
        \includegraphics[width=0.85\columnwidth]{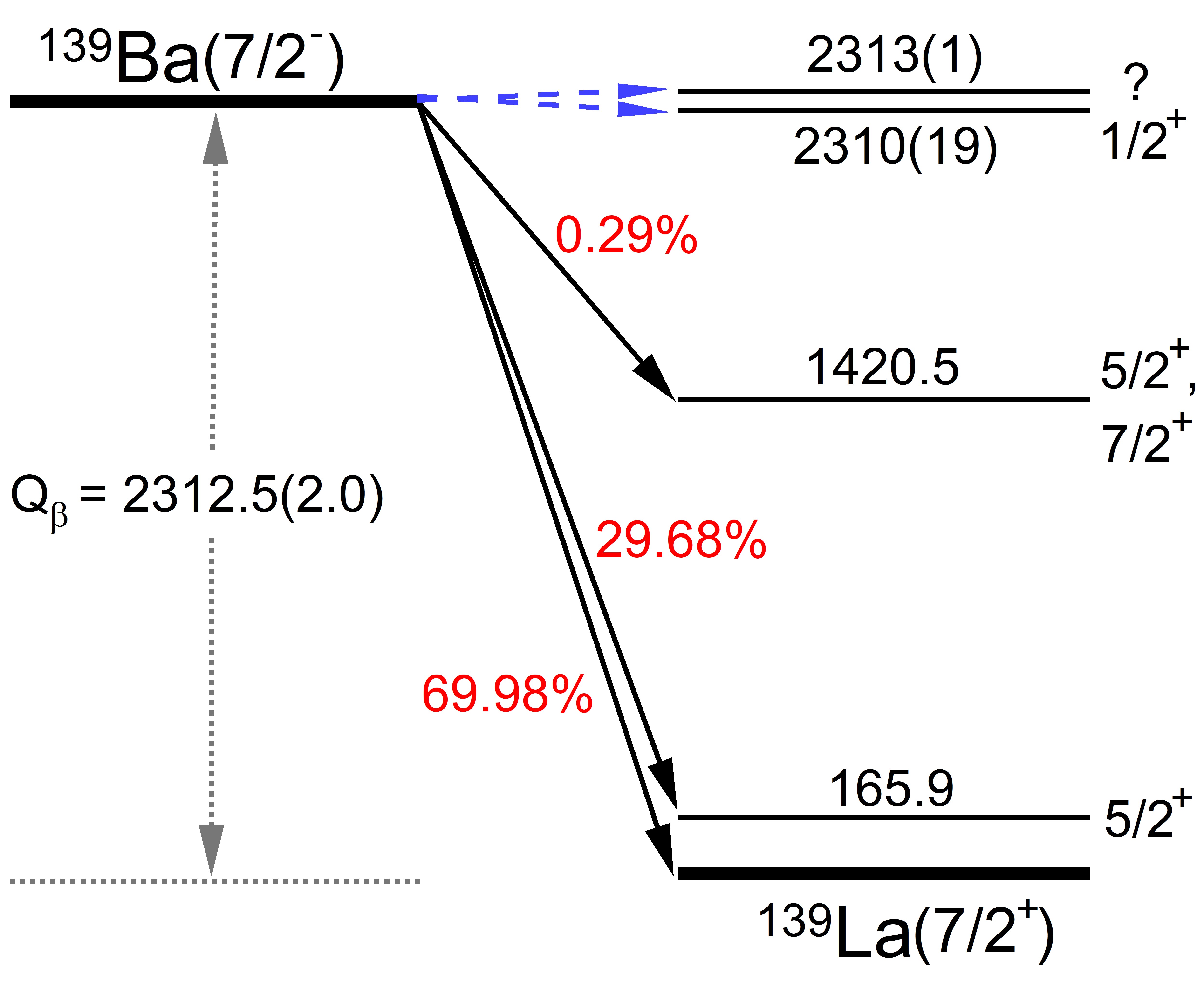}
    \caption{(color online) Decay schemes for $^{89}$Sr and $^{139}$Ba showing the main $\beta$-decay branches (solid black arrows) and the potential ultra-low $Q$-value decay branches (dashed blue arrows) investigated in this work. The ground-state to ground-state $Q$-values are obtained using data from the AME2016~\cite{Wan2017}. All values are given in units of keV.}
    \label{fig:levels}
\end{figure}

In Fig.~\ref{fig:levels}, decay schemes are shown for $^{89}$Sr and $^{139}$Ba, with the main $\beta$-decay transitions indicated by solid black arrows and the potential ultra-low $Q$-value decays indicated by dashed blue arrows. In the case of $^{89}$Sr, the potential ultra-low $Q$-value decay is to the 3/2$^{-}$ state in $^{89}$Y at 1507.4 keV. For $^{139}$Ba, there are two potential ultra-low $Q$-value decay branches: to the 1/2$^{-}$ state in $^{139}$La at 2310 keV and to the state of unknown spin and parity at 2313 keV. The ground-state to ground-state $Q$-values given in Fig.~\ref{fig:levels} are calculated using data from the most recent atomic mass evaluation, AME2016~\cite{Wan2017}, and are limited by the 1.6 keV/c$^{2}$ and 2.0 keV/c$^{2}$ uncertainties in the masses of $^{89}$Y and $^{139}$La, respectively. The mass of the parent isotopes, $^{89}$Sr and $^{139}$Ba, are known to 0.09 keV/c$^{2}$ and 0.32 keV/c$^{2}$, respectively. Hence, precise and accurate atomic masses for $^{89}$Y and $^{139}$La with uncertainties $<$1 keV/c$^{2}$ are called for to determine if these potential ultra-low $Q$-value decay branches are energetically allowed. In this paper we present the first direct mass measurements of $^{89}$Y and $^{139}$La using Penning trap mass spectrometry. We calculate new $Q$-values for these decays and discuss implications for potential 
ultra-low $Q$-value $\beta$-decays in $^{89}$Sr and $^{139}$Ba.

\section{Experiment Description}
The atomic masses of $^{89}$Y and $^{139}$La were measured at the Low Energy Beam and Ion Trap (LEBIT) facility, located at the National Superconducting Cyclotron Laboratory (NSCL)~\cite{rin2013}. While LEBIT was designed to perform on-line mass measurements of rare isotopes from the NSCL produced via projectile fragmentation, it also houses a Laser Ablation Source (LAS)~\cite{izz2016} and a Thermal Ion Source (TIS) which can be used to produce stable and long-lived isotopes for use as reference masses and for offline measurements with applications in neutrino and nuclear physics~\cite{Red2012,Bus2013,Bus2013b,Lin2013,gul2015,eib2016,gam2016,kan2017,san2019}. For the $^{139}$La measurement, the LAS was fitted with a 25mm $\times$ 25mm $\times$ 1mm thick sheet of lanthanum~\cite{espi}, used to produce $^{139}$La$^+$ (99.9\% natural abundance). The TIS was fitted with a canister of xenon gas to produce $^{136}$Xe$^+$ (8.9\% natural abundance) via plasma ionization for use as a reference ion. For the $^{89}$Y measurement, the LAS was fitted with a 25mm $\times$ 25mm $\times$ 1mm thick sheet of yttrium~\cite{espi}, used to produce $^{89}$Y$^+$ (100\% natural abundance), and the TIS was set up to produce $^{85}$Rb$^+$ and $^{87}$Rb$^+$ (72.2\% and 27.8\% natural abundances, respectively) via surface ionization for use as reference ions.

\begin{figure}[h]
\includegraphics[width=\columnwidth]{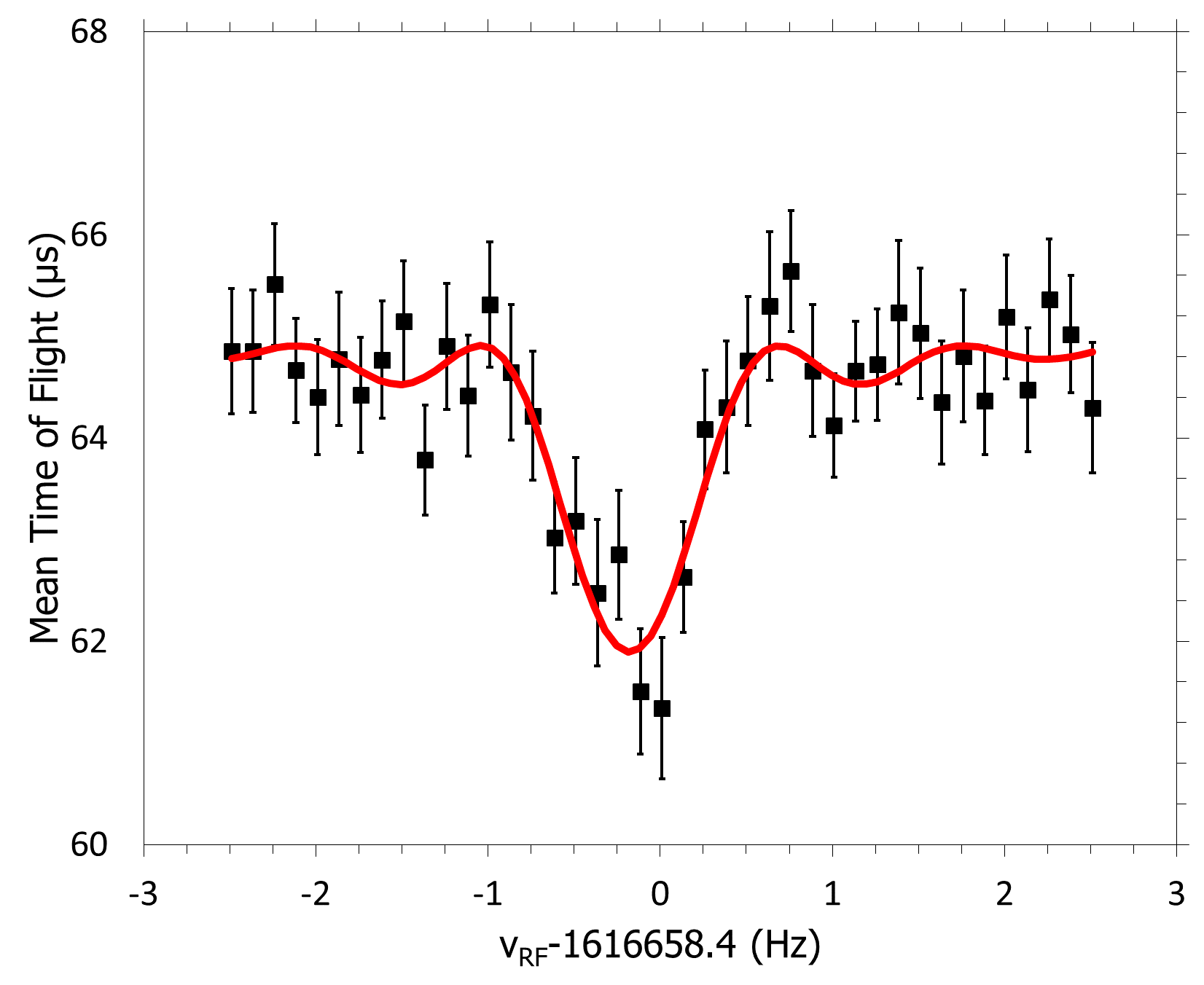}
\caption{(Color online) A $t_{rf} =1$ s cyclotron frequency resonance curve for $^{89}$Y (see text for details). The solid red line is a fit to the theoretical line shape~\cite{kon1995}.\label{fig:Resonance}} 
\end{figure}

The LEBIT Penning trap is a hyperbolic trap housed in a 9.4 T magnetic field. The facility uses the Time of Flight-Ion Cyclotron Resonance (TOF-ICR) technique to precisely measure the cyclotron frequency of the ion in question~\cite{Gra1980}. Ions held within the trap are driven with a quadrupolar radio frequency (rf) pulse near to the cyclotron frequency for a period of time, $t_{rf}$. They are released towards a micro-channel plate (MCP) detector and the time-of-flight between the trap and the detector is measured. The time-of-flight is minimized when the frequency of the rf pulse matches the cyclotron frequency of the ion in question. By varying the frequency of the rf pulses around the cyclotron frequency and taking multiple time-of-flight measurements, a resonance curve can be built and fit to a theoretical line shape (see Fig.~\ref{fig:Resonance}). The width of the resonance, and hence the precision to which the central frequency can be obtained from a fit to the theoretical line shape, goes as $\sim1/t_{rf}$. In this work we used $t_{rf} = 1 $s. Before and after each measurement of the ion of interest, a cyclotron frequency measurement is taken with the reference ion. The reference measurements are linearly interpolated to find the cyclotron frequency of the reference ion at the time of the measurement of the ion of interest. 

The cyclotron frequency of an ion with a charge-to-mass ratio of $q/m$ is given by the relationship

\begin{equation}
f_c = \frac{qB}{2\pi m}.
\end{equation}

From the cyclotron frequency of the reference ion and the ion of interest, the cyclotron frequency ratio, corresponding to the inverse mass ratio of the ions can be obtained:

\begin{equation}
R = \frac{f_c^{int}}{f_c^{ref}} = \frac{m_{ref}}{m_{int}}.
\end{equation}

A series of measurements of $R$ are taken to find an average value, $\bar{R}$. The atomic mass can then be obtained using the known mass of the reference atom and the equation
\begin{equation}
M_{int} = (M_{ref}-m_e)\frac{1}{\bar{R}}+m_e,
\end{equation}
where $M_{int}$ is the atomic mass of the atom of interest, $M_{ref}$ is the atomic mass of a well-known reference atom, and $m_e$ is the mass of the electron. We have ignored the binding energy of electrons in singly charged ions as they are $\lesssim$10 eV, which is much smaller than our statistical uncertainty and therefore negligible. The calculated daughter mass can then be used with the mass of the parent atom to find the $Q$-value of the ground-state to ground-state decay, using the equation
\begin{equation}
Q = (M_p-M_d)c^2, \label{eq:Q}
\end{equation}
where $M_d$  is the atomic mass of the daughter (corresponding to $M_{int}$ for $^{89}$Y and $^{139}$La measured here) and $M_p$ is the atomic mass of the parent ($^{89}$Sr and $^{139}$Ba taken from the AME2016).

\section{Results and Discussion}
The average cyclotron frequency ratios, $\bar{R}$, can be found in Table~\ref{table:ratio}. These ratios have had small corrections applied to them to correct for the $\Delta R$/$\Delta m$ = 2 $\times$ 10$^{-10}$ per u shift to the ratio that occurs in our system when measuring non-mass doublets~\cite{gul2015}. The uncertainties have been inflated by the Birge Ratio~\cite{Bir1932} to allow for potential systematic uncertainties that may not have been accounted for.

\begin{table}[]
\caption{\label{table:ratio} Measured cyclotron frequency ratios for $^{89}$Y$^{+}$ and $^{139}$La$^{+}$ ions against their reference ions. N is the number of individual ratio measurements contributing to the average, $\bar{R}$. The uncertainties for $\bar{R}$, shown in parentheses, have been inflated by the Birge Ratio, BR, when BR $>$ 1.}
\begin{ruledtabular}
\begin{tabular}{ccccc}
Num. & Ion Pair & N & BR & $\bar{R}$ \\
\hline
(i) & $^{89}\text{Y}^+$/$^{87}\text{Rb}^+$ & 66 & 1.2 & $0.977\ 541\ 739\ 2(56)$ \\
(ii) & $^{89}\text{Y}^+$/$^{85}\text{Rb}^+$ & 44 & 1.1 & $0.955\ 075\ 250\ 9(56)$  \\
(iii) & $^{139}\text{La}^+$/$^{136}\text{Xe}^+$ & 66 & 1.3 & $0.978\ 408\ 760\ 7(47)$ \\
\end{tabular}
\end{ruledtabular}
\end{table}

The ratios in Table~\ref{table:ratio} were used to obtain absolute atomic masses for $^{89}$Y and $^{139}$La. The mass excesses were then calculated using the equation
\begin{equation}
ME = (M_{int}-A)\times 931\,494.0954(57) \rm{(keV/c^2)/u},
\end{equation}
where $A$ is the mass number of the atom of interest and the conversion factor is from Ref.~\cite{Moh2014}. The results are listed in Table~\ref{table:mass} and are compared with the values from the AME2016~\cite{Wan2017}. The mass differences are also shown in Fig. 3. There is a 2.8 keV/c$^{2}$ reduction in the $^{89}$Y mass excess obtained in this work compared to the AME2016. In the AME, the $^{89}$Y mass value was obtained mainly from a neutron capture measurement linking it to $^{90}$Y, which is then linked to $^{90}$Zr through a $^{90}$Y $\beta$-decay measurement. The mass of $^{90}$Zr was measured directly at LEBIT~\cite{gul2015}.

In the case of $^{139}$La there is a 4.0 keV/c$^{2}$ increase in the mass excess value from this work compared to the AME2016. The mass value of $^{139}$La in the AME is not based on a direct measurement, but through a $\beta$-decay measurement that links it to the mass of $^{139}$Ba and through a network of neutron capture, $\beta$-decay, and $\alpha$-decay measurements which eventually link it to $^{163}$Dy and $^{163}$Ho, for which precise Penning trap measurements have been performed~\cite{eli2015}. In a previous measurement campaign we performed a direct measurement of the mass of $^{138}$La and found a +5.8 keV/c$^{2}$ discrepancy compared to the AME2016 \cite{san2019}. $^{138}$La was determined in the AME2016 mainly via a $^{138}$La$(d,p)^{139}$La reaction measurement with an uncertainty of $\sim$3 keV. Hence, our results for the two lanthanum isotopes are consistent with the $^{138}$La$(d,p)^{139}$La measurement being correct, and the $^{139}$La mass in AME2016 being off by 4 keV/c$^{2}$.

\begin{table}[]
\caption{\label{table:mass} Mass excesses for $^{89}$Y and $^{139}$La obtained in the work along with results from the AME2016~\cite{Wan2017} and the difference $\Delta$ME = ME$_{\rm{LEBIT}}$ - ME$_{\rm{AME}}$}
\begin{ruledtabular}
\begin{tabular}{ccccc}
\multirow{2}{*}{Isotope} & \multirow{2}{*}{Ref.} & \multicolumn{1}{c}{This work} & \multicolumn{1}{c}{AME2016} & \multicolumn{1}{c}{$\Delta$ME}\\
 &  & \multicolumn{1}{c}{(keV/c$^2$)} & \multicolumn{1}{c}{(keV/c$^2$)} & \multicolumn{1}{c}{(keV/c$^2$)}\\
\hline
\multirow{3}{*}{$^{89}$Y} & $^{87}$Rb & $-87\ 710.67(0.47)$ & \multirow{3}{*}{$-87\ 708.4(1.6)$} & -2.3(1.7)\\
 & $^{85}$Rb & $-87\ 711.78(0.49)$  &   & -3.4(1.7)\\
 & Ave. & $-87\ 711.21(0.34)$ & & -2.8(1.6)\\
\hline
$^{139}$La & $^{136}$Xe & $-87\ 222.15(0.62)$ & $-87\ 226.2(2.0)$ & 4.0(2.1)\\
\end{tabular}
\end{ruledtabular}
\end{table}

\begin{figure}[]
\includegraphics[width=\columnwidth]{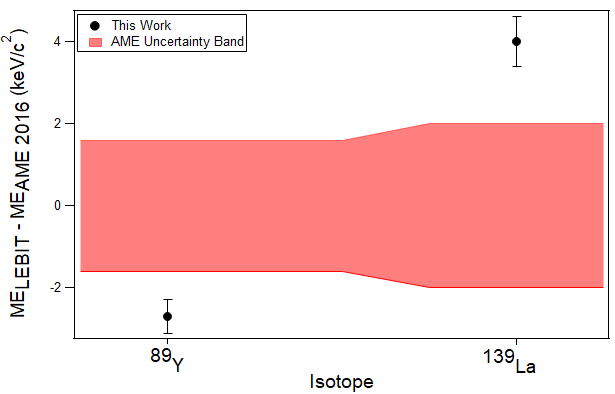}
\caption{The mass excesses measured in this work. The red bands show the AME2016 uncertainty and the black dots are the measured values. (Color online) \label{fig:QValues}} 
\end{figure}

Using our new atomic masses for $^{89}$Y and $^{139}$La along with masses for $^{89}$Sr and $^{139}$Ba from AME2016, we obtain new ground-state to ground-state $Q$-values, $Q_{GS}$, which are listed in Table~\ref{table:Q}. We also list the energy of the potential ultra-low $Q$-value decay daughter state and the calculated $Q$-value for the ultra-low decay branch from
\begin{equation}
  Q_{UL} = Q_{GS} - E^{*}.    
\end{equation}
For the decay of $^{89}$Sr to $^{89}$Y, the $Q$-value was increased by 2.8 keV. The new value of 1502.20 keV is still less than the  $^{89}$Y 3/2$^-$ excited state energy of 1507.4 keV. With $Q_{UL} = -5.20(0.37)$ keV, it can now be said definitively that the 3/2$^-$ excited state is not a candidate for ultra-low $Q$-value decay. We note that the mass of $^{89}$Sr is known to 0.09 keV/c$^{2}$ via an ($n,\gamma)$ measurement that links it to $^{88}$Sr, which has been measured precisely using Penning trap mass spectrometry~\cite{san2019}.

For the decay of $^{139}$Ba to $^{139}$La, the $Q$-value was decreased by 4 keV. The new value of 2308.37 keV is now substantially less than the 2313 keV excited state of $^{139}$La. With $Q_{UL} = -4.6(1.2)$ keV, it can now be said definitively that the 2313 keV excited state is not energetically viable for ultra-low $Q$-value decay. However, the 1/2$^+$ excited state of $^{139}$La, with an energy of 2310(19) keV and $Q_{UL} = -1.6(19.0)$ keV, still has too large of an uncertainty for any definitive claims to be made. The energy of this excited state will need to be measured to a higher precision to determine if it is a candidate for an ultra-low $Q$-value $\beta$-decay. The mass of $^{139}$Ba is known to 0.32 keV/c$^{2}$ via an $(n,\gamma)$ measurement linking it to $^{138}$Ba. In Ref.~\cite{san2019} we also performed a direct measurement of the mass of $^{138}$Ba, which was in excellent agreement with the AME2016 result. The mass of $^{138}$Ba was derived in the AME2016 from an $(n,\gamma)$ measurement linking it to $^{137}$Ba---the same series of measurements linking $^{138}$Ba and $^{139}$Ba. This chain of measurements is ultimately anchored to $^{136}$Xe and $^{133}$Cs, for which precise atomic mass measurements have been performed. Hence, there is good reason to accept the AME2016 $^{139}$Ba mass.

\begin{table}[h]
 \caption{\label{table:Q} $Q$-values based on the absolute mass measurements in Table~\ref{table:mass} and Eqn.~\ref{eq:Q}. The column $E^*$ gives the energy of the excited state of the daughter nucleus. The result for the ultra-low $Q$-value decay branch is calculated as $Q_{UL}$ = $Q_{GS} - E^*$.}
\begin{ruledtabular}
\begin{tabular}{ccccc}
\multirow{2}{*}{Parent} & \multirow{2}{*}{Daughter} & \multicolumn{1}{c}{$Q_{GS}$} & \multicolumn{1}{c}{E*} & \multicolumn{1}{c}{$Q_{UL}$} \\
 & & \multicolumn{1}{c}{keV} & \multicolumn{1}{c}{keV} & \multicolumn{1}{c}{keV} \\
\hline
$^{89}$Sr & $^{89}$Y & 1502.20(0.35) & 1507.4(0.1) & -5.20(0.37) \\
$^{139}$Ba & $^{139}$La & 2308.37(0.68) & 2310(19) & -1.6(19.0) \\
$^{139}$Ba & $^{139}$La & 2308.37(0.68) & 2313(1) & -4.6(1.2)\\
\end{tabular}
\end{ruledtabular}
\end{table}

\section{Conclusion}

Using Penning trap mass spectrometry, the mass excess of $^{89}$Y was measured to be $-87711.21(0.34)$ keV/c$^{2}$ and the mass excess of $^{139}$La was measured to be $-87222.15(0.62)$ keV/c$^{2}$. These are the first Penning trap mass spectrometry measurements of either isotope. The new masses were used to calculate the $\beta$-decay $Q$-values for $^{89}$Sr $\rightarrow ^{89}$Y and $^{139}$Ba $\rightarrow ^{139}$La. The $Q$-value for $^{89}$Sr was found to be 1502.20(0.35) keV and the $Q$-value for $^{139}$Ba was found to be 2308.37(0.68) keV. Both have had their uncertainties reduced by more than a factor of two. For the decay of $^{89}$Sr, the potential ultra-low $Q$-value decay channel to the 3/2$^-$ state in $^{89}$Y at 1507.4 keV has been refuted. For the decay of $^{139}$Ba, one potential ultra-low $Q$-value decay channel to the 2313 keV level  in $^{139}$La with unknown $J^{\pi}$ has been refuted. However, the 1/2$^+$ excited state in$^{139}$La, currently measured to be 2310(19)keV, is still a candidate. More precise measurements of the excitation energy of $^{139}$La will be necessary to determine whether or not the $\beta$-decay of $^{139}$Ba to this state is an ultra-low $Q$-value decay candidate. 

\section*{Acknowledgments}

This research was supported by Michigan State University and the Facility for Rare Isotope Beams and the National Science Foundation under Contracts No. PHY-1102511 and No. PHY1307233. This material is based upon work supported by the US Department of Energy, Office of Science, Office of Nuclear Physics under Award No. DE-SC0015927.   

\bibliography{ULQV_paper.bib} 

\end{document}